\documentclass[10pt,conference]{IEEEtran}

\usepackage[hidelinks]{hyperref}
\usepackage{cite}
\usepackage{tabularx}
\usepackage{booktabs}
\usepackage{multirow}
\usepackage{threeparttable}
\usepackage{xspace}
\usepackage[table]{xcolor}
\usepackage{graphicx}
\usepackage{listings}
\usepackage{tcolorbox}
\usepackage{siunitx}

\newcolumntype{Y}{>{\raggedright\arraybackslash}X}
\newcommand{\TOOL}{LintQ-LLM\xspace}

\lstdefinestyle{prompt}{frame=tb,
	showstringspaces=false,
	columns=flexible,
	basicstyle={\scriptsize\ttfamily},
	numbers=left,
	xleftmargin=2.5em,
	breaklines=true,
	breakatwhitespace=true,
	tabsize=2,    
    moredelim=[s][\color{purple}]{?}{?},
}

\begin{document}

\title{Quantum Program Linting with LLMs: Emerging Results from a Comparative Study
}

\author{\IEEEauthorblockN{Seung Yeob Shin}
\IEEEauthorblockA{\textit{University of Luxembourg} \\
Luxembourg, Luxembourg \\
seungyeob.shin@uni.lu}
\and
\IEEEauthorblockN{Fabrizio Pastore}
\IEEEauthorblockA{\textit{University of Luxembourg} \\
Luxembourg, Luxembourg \\
fabrizio.pastore@uni.lu}
\and
\IEEEauthorblockN{Domenico Bianculli}
\IEEEauthorblockA{\textit{University of Luxembourg} \\
Luxembourg, Luxembourg \\
domenico.bianculli@uni.lu}
}

\maketitle

\begin{abstract}
Ensuring the quality of quantum programs is increasingly important; however, traditional static analysis techniques are insufficient due to the unique characteristics of quantum computing.
Quantum-specific linting tools, such as LintQ, have been developed to detect quantum-specific programming problems; however,  they typically rely on manually crafted analysis queries.
The manual effort required to update these tools limits their adaptability to evolving quantum programming practices.

To address this challenge, this study investigates the feasibility of employing Large Language Models (LLMs) to develop a novel linting technique for quantum software development and explores potential avenues to advance linting approaches.
We introduce \TOOL, an LLM-based linting tool designed to detect quantum-specific problems comparable to those identified by LintQ.
Through an empirical comparative study using real-world Qiskit programs, our results show that \TOOL is a viable solution that complements LintQ, with particular strengths in problem localization, explanation clarity, and adaptability potential for emerging quantum programming frameworks, thus providing a basis for further research.
Furthermore, this study discusses several research opportunities for developing more advanced, adaptable, and feedback-aware quantum software quality assurance methods by leveraging LLMs.
\end{abstract}

 \section{Introduction}
\label{sec:intro}

Quantum computing has made notable advancements in recent years, offering the potential to efficiently solve a certain set of problems in various domains, such as chemistry~\cite{aspuru2005,mcardle2020}, cryptography~\cite{ekert1991,portmann2022}, and optimization~\cite{harrow2009,ebadi2022q}.
As the field of quantum computing progresses, the quality assurance of quantum programs becomes increasingly important.
Quantum programs manipulate both quantum and classical bits, where quantum bits (i.e., qubits) follow the principles of quantum mechanics, such as superposition and entanglement.
Due to the unique characteristics of quantum programs, traditional software analysis techniques are insufficient for ensuring the quality of quantum programs.
Hence, specialized analysis techniques that account for the specificities of quantum programs are needed.

Static program analysis techniques~\cite{Nielson1999,Dunsmuir1985}, particularly linting techniques, have been successfully applied to the development of classical software programs.
Linting techniques automatically detect potential (or definitive) problems, including violations of coding rules, deviations from best practices, or defects.
Linting tools scan the source code under analysis, matching it against a predefined set of rules to detect potential problems.
These tools can warn developers about problems before they execute the code, aiming at improving software quality.
In the quantum domain, linting tools play an even more important role, as executing quantum programs is costly and often restricted by limited access to quantum computers.

To account for the unique characteristics of quantum programs during linting, some tools have been introduced recently, including QSmell~\cite{ChenCCSA23}, QChecker~\cite{ZhaoWLZ23}, QCPG~\cite{KaulKB23}, and LintQ~\cite{PaltenghiP24}.
Among them, LintQ is the state-of-the-art tool.
It targets Qiskit programs~\cite{Qiskit} and
implements a set of analyses to detect quantum-specific programming problems, such as those related to measurements, gate usage, resource allocation, and implicit constraint violations in API usage.

While quantum-specific linting tools have demonstrated their feasibility and effectiveness, they rely on manually crafted patterns, rules, or queries, which limit their adaptability to evolving quantum programming practices.
As the field of quantum computing advances, new libraries, frameworks, and programming paradigms are expected to emerge, making it increasingly challenging to manually maintain and update these linting tools.

To address the above-mentioned challenge, this study investigates the feasibility of using Large Language Models (LLMs) to detect quality problems in quantum programs.
Given the remarkable success of LLMs in assisting programming activities in classical software quality assurance~\cite{LLMtestingSurvey}, we aim to assess whether LLMs could offer a promising alternative or complement existing linting tools.
Our hypothesis is that LLM can leverage their extensive knowledge of programming patterns, best practices, and implicit coding conventions, obtained from vast amounts of code repositories, including those related to quantum programs.
In addition, LLMs have the potential to provide contextualized information and suggestions tailored to the quantum source code under analysis.
Furthermore, LLM-enabled interactive, conversational analysis---which leverages chatbot-style feedback loops during the implementation of quantum programs---may enhance the linting process by offering explanations and recommendations in a more intuitive and user-friendly manner, thereby lowering the entry barrier for developers in developing high-quality quantum software programs.

Given the potential advantages of LLMs in enhancing the linting process, we first investigate the feasibility of an LLM-based linting approach in comparison to the state-of-the-art quantum-specific linting tool, LintQ.
In addition, we explore how LLMs can complement and extend existing linting techniques, providing a more adaptable, accurate, and developer-friendly approach to quantum software quality assurance.

To summarize, this NIER paper makes the following contributions:
(1)~We introduce \TOOL, an LLM-based linting tool that performs analyses comparable to those of LintQ.
(2)~We conduct a comparative study of \TOOL and LintQ, empirically evaluating both tools on the same dataset of real-world Qiskit programs.
(3)~We analyze the strengths and weaknesses of these tools and, based on our findings, propose future research directions.
To our knowledge, this is the first attempt to leverage LLMs for automatically detecting quantum programming problems.
We believe this work serves as a stepping stone toward more advanced LLM-powered quantum software quality assurance tools that provide adaptable and precise analyses along with context-aware explanations and suggestions.

The rest of the paper is organized as follows.
Section~\ref{sec:background} provides background on LintQ.
Section~\ref{sec:approach} describes our approach to developing \TOOL.
Section~\ref{sec:exp} presents our comparative experimental results.
Section~\ref{sec:discussion} discusses the findings from our experiments and outlines directions for future work.
Section~\ref{sec:related} surveys related work.
Section~\ref{sec:discussion}  concludes the paper. \section{Background: LintQ}
\label{sec:background}

LintQ~\cite{PaltenghiP24} is a static analysis framework that detects quantum-specific problems in Qiskit source code.
Specifically, LintQ introduces a set of abstractions for common quantum concepts, such as quantum registers, classical registers, quantum circuits, gates, qubit usage, and measurements. These abstractions enable LintQ to perform static analyses of Qiskit source code. Compared to other tools such as QChecker~\cite{ZhaoWLZ23} and QSmell~\cite{ChenCCSA23}, these abstractions enable LintQ to perform static analyses quickly, without the need for processing the underlying implementation.
Building upon these abstractions, LintQ provides ten analyses, each of which identifies a quantum-specific programming problem.
For the analyses, LintQ leverages CodeQL~\cite{AvgustinovMJS16}, a general-purpose static analysis engine for source code.
Each analysis is constructed as a query supported by CodeQL over the behavioral representation of the Qiskit code under analysis, which is expressed using the abstractions.

{
\renewcommand{\arraystretch}{0.9}
\begin{table}[t]
\small
\begin{threeparttable}
\caption{Problems identified by LintQ and their descriptions.}
\label{tab:lintq problems}
\centering
\begin{tabularx}{\columnwidth}{l@{\hspace{.5em}}Y}
\toprule
\textbf{Problem}\tnote{a} & \textbf{Description} \\
\midrule
\rowcolor{gray!30}\multicolumn{2}{c}{\textbf{Measurement- or Gate-related Problems}} \\
DoubleMeas & Two consecutive measurements are performed on the same qubit state. \\
OpAfterMeas & A gate is applied to a qubit after it has already been measured. \\
MeasAllAbuse & Measurement results are stored in a newly and implicitly created register, despite the presence of an existing classical register. \\
CondWoMeas & A conditional gate is applied without measuring the associated register. \\
ConstClasBit & A qubit is measured without undergoing any prior transformation. \\
\rowcolor{gray!30}\multicolumn{2}{c}{\textbf{Resource Allocation Problems}} \\
InsuffClasReg & There are not enough classical bits to store the measurement results of all qubits. \\
OversizedCircuit & The quantum register includes qubits that remain unused. \\
GhostCompose & Two circuits are composed, but the resulting composed circuit is not utilized. \\
\rowcolor{gray!30}\multicolumn{2}{c}{\textbf{Implicit API Constraint Violations}} \\
OpAfterOpt & A gate is applied to the circuit after transpilation. \\
OldIdenGate & An identity gate is created using an API that has been removed. \\
\bottomrule
\end{tabularx}
\begin{tablenotes}
\item[a] {\footnotesize Note that the problem names match those used in the LintQ paper~\cite{PaltenghiP24} for each analysis.}
\end{tablenotes}
\end{threeparttable}
\end{table}
}

Table~\ref{tab:lintq problems} presents the quantum-specific programming problems that LintQ identifies in Qiskit source code.
Each problem corresponds to an invalid or undesirable use of Qiskit programming constructs that may arise during quantum program development.
The ten problems listed in Table~\ref{tab:lintq problems} are categorized into three groups based on their nature: (1)~measurement- or gate-related problems, (2)~resource allocation problems, and (3)~implicit API constraint violations.

LintQ was applied to 7,568 real-world Qiskit programs and achieved an overall precision of 62.5\%, which was computed by manually inspecting 361 warnings generated by the tool and observing 261 correctly reported problems.
Further, its application to a benchmark of 42 quantum programs led to a recall of 7.1\%, which shows large room for improvement despite LintQ being the state-of-the-art approach.
 \section{Approach: LLM-based Linter}
\label{sec:approach}

This section describes \TOOL, our LLM-based linting approach for quantum programs.
Our implementation of \TOOL is available online~\cite{REPLICABILITY}.

\subsection{Overview}
\label{subsec:overview}

\begin{figure}[t]
\centering
\includegraphics[width=0.82\columnwidth]{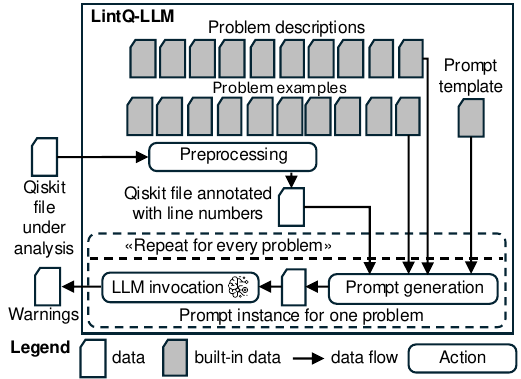}
\caption{An overview of the data flow in \TOOL.}
\label{fig:overview}
\end{figure}

Figure~\ref{fig:overview} provides an overview of \TOOL. \TOOL takes a Qiskit source code file as input and produces warnings for any quantum-specific problems identified, including their specific locations and explanations.
Its key characteristics are to process one source code file a time, which is common for other linters, and more importantly, querying the LLM independently for each type of problem to be identified in the source code.
An alternative would have been to query the LLM all at once, for all the potential problems to be identified, but this would have led to longer prompts with a lot of instructions for the LLM.
Such long prompts reduce the number of tokens available for the program under analysis; indeed, an LLM can process only a fixed number of tokens (for simplicity, characters), thus reducing the maximum length of the file under analysis, which is provided within the prompt, as described below.
Further, querying the LLM for multiple problems a time is more likely to introduce mistakes in the generated answer due to the increased length of the prompt and potential ambiguities among the multiple instructions.

\subsection{Prompt Engineering}
\label{subsec:prompt engineering}

We created LLM prompts aimed at detecting quantum-specific programming problems.
Specifically, the prompts targeted the ten problems listed in Table~\ref{tab:lintq problems}, ensuring a fair comparison between LintQ and \TOOL.
We note that the first arXiv version of the LintQ paper~\cite{paltenghi2023}
was posted on 1 October 2023, and the initial commit on the LintQ repository~\cite{PaltenghiLintQ}
was made on 26 October 2022.
To avoid potential biases or confounding factors resulting from LLMs having access to publicly available data about LintQ, we selected the GPT-3.5 Turbo model~\cite{OpenAIgpt35}
from OpenAI, whose knowledge cutoff date is 1 September 2021.

To detect a particular quantum-specific problem in the source code file, \TOOL generates a dedicated prompt for that problem, which is then fed to the LLM. To this end, \TOOL uses a prompt template, which is common for all the types of problems addressed by \TOOL and is instantiated with appropriate data at each invocation.

\begin{figure}[t]
\begin{lstlisting}[style=prompt,basicstyle=\fontsize{6}{7.2}\ttfamily]
## Situation
You are analyzing the source code to detect ?problem? occurrences.
<code>
?code?
</code>

## Your Role
Act as a source code linter tool to detect all occurrences of the problem:
?problem_description?

## Output Format
If the code contains ?problem?, return this JSON object:
{
	"problem": "?problem?", 
	"snippets": ["string"], // extract code verbatim where ?problem? occurs.
	"lines": ["integer"], // list line numbers where ?problem? occurs.
	"explanations": ["string"] // explain why each line has ?problem?.
}
\end{lstlisting}
\caption{Prompt template. \textcolor{purple}{\texttt{?param?}} indicates a parameter in the template.}
\label{fig:template}
\end{figure}

Figure~\ref{fig:template} shows a shortened version of our template;
the complete prompt template is available online~\cite{REPLICABILITY}.
As shown in Figure~\ref{fig:template}, we structured the prompt template based on prompt engineering guidelines~\cite{burkov2025,promptingguide} and leveraged few-shot learning~\cite{BrownMRSKDNSSAA20}.
The template includes three parameters that shape the concrete prompts, as follows:
\texttt{(?problem?)} the name of the quantum-specific problem to detect, i.e., one of the problems listed in Table~\ref{tab:lintq problems},
\texttt{(?code?)} the source code to be analyzed for detecting occurrences of the problem,
\texttt{(?problem\_description?)} a detailed description of the problem, including example cases for few-shot learning. Before querying the LLM, these parameters are programmatically replaced with the corresponding data.
For \texttt{(?code?)}, we provide the source code under analysis verbatim, but we annotate each line with the corresponding line number, which we found to be necessary to avoid erroneous outputs from the LLM.
An example problem description is provided in the following paragraphs, \hbox{after an overview of the template structure.}

In the \texttt{Situation} section of the template (Figure~\ref{fig:template}), the context in which the LLM operates is explicitly defined, ensuring the LLM understands its objective of identifying a specific quantum-related programming problem (provided by \texttt{?problem?}) in the provided source code (given by \texttt{?code?}).
The \texttt{Your Role} section instructs the LLM to act as a source code linter, specifically focusing on detecting the problem described by \texttt{?problem\_description?}.
The \texttt{Output Format} section enforces a strict JSON-based output format to facilitate automated parsing and analysis.
The output includes: (\texttt{problem}), i.e.,  the name of the detected quantum-specific problem; (\texttt{snippets}), i.e., extracted code segments where the problem occurs; (\texttt{lines}), i.e.,  the line numbers corresponding to the instructions with the identified problem;
(\texttt{explanations}), i.e., descriptions of why the reported code segments exhibit the problem.

\begin{figure}[t]
\begin{lstlisting}[style=prompt,basicstyle=\fontsize{6}{7.2}\ttfamily]
DoubleMeas (Double measurement) --- Any two subsequent measurements on the same qubit produce the same classical result, making the second measurement not only redundant but also a possible sign of unintended behavior or a misunderstanding of the properties of quantum information.
The code example below shows the DoubleMeas problem.
<example>
	circuit = QuantumCircuit(3, 3)
	circuit.ccx(0, 1, 2)
	circuit.measure(0, 0) # Measure qubit 0
	circuit.measure(2, 2)
	circuit.measure(0, 1) # Problem: Qubit 0 already measured
</example>
\end{lstlisting}
\caption{Prompt description of the double measurement problem (DoubleMeas in Table~\ref{tab:lintq problems}). We reuse the problem description from the LintQ paper~\cite{PaltenghiP24}.
}
\label{fig:problem desc}
\end{figure}

As an example, Figure~\ref{fig:problem desc} shows the prompt description for the double measurement problem (DoubleMeas in Table~\ref{tab:lintq problems}).
The description explains how consecutive measurements on the same qubit always yield the same classical result, making the second measurement redundant and potentially indicating unintended behavior or a misunderstanding of quantum information principles.
The \texttt{<example>} block in the description presents the DoubleMeas problem in a Qiskit circuit.
In the circuit, qubit 0 is measured twice on lines 6 and 8, with comments explaining the problem.
We reuse the problem descriptions from the LintQ paper~\cite{PaltenghiP24} 
to ensure consistency with the LintQ work, allowing for comparability in our evaluation.
However, \TOOL does not use any descriptions of how LintQ identifies quantum-specific problems, ensuring that \TOOL relies solely on the LLM's analysis capabilities for a fair comparison with LintQ.
 \section{Experiments: A comparative study}
\label{sec:exp}

In this section, we report on our experiments comparing LintQ and \TOOL. Specifically, we aim to address the following research question (RQ): \emph{How does \TOOL compare to LintQ in terms of effectiveness in detecting quantum-specific problems in Qiskit code?}

\subsection{Datasets and Metrics}
\label{subsec:datasets}

We used the annotated dataset from the LintQ study~\cite{PaltenghiLintQ}
to enable a direct comparison with LintQ and to avoid introducing threats to internal validity.
The dataset includes 345 warnings, randomly sampled by LintQ authors from the LintQ-identified warnings for the 7,568 Qiskit files processed in their experiments.
The LintQ authors annotated these warnings as true positives (TPs), false positives (FPs), and noteworthy (NWs).
TPs are warnings that correctly identify real quantum-specific problems in the analyzed source code.
FPs refer to warnings that do not correspond to 
actual problems in the Qiskit code.
NWs are potential problems for which the authors were unable to definitively determine whether the behaviors caused by such problems were intended by developers or not.

The 345 warnings in the dataset belong to 268 Qiskit files.
Since \TOOL uses ten example cases for few-shot learning (six from the Qiskit files and four from the LintQ paper)
 we excluded the annotated warnings of the six files from our evaluation dataset.
As a result, our evaluation dataset contains 338 annotated warnings belonging to 262 Qiskit files.

For our experiment, we applied \TOOL to each file in the evaluation dataset
and measured effectiveness in terms of precision and recall.
Precision is the number $\mathit{tp}$ of TPs identified by \TOOL divided by the number of warnings that overlap between the evaluation dataset and \TOOL's detected warnings.
These overlapping warnings contain $\mathit{tp}$, $\mathit{fp}$, and $\mathit{nw}$ instances of TPs, FPs, and NWs; hence, precision is calculated as $\frac{\mathit{tp}}{ (\mathit{tp} + \mathit{fp} + \mathit{nw})}$.
Recall is the number $\mathit{tp}$ of TPs identified by \TOOL divided by the total number of TPs in the evaluation dataset.
This total includes both the $\mathit{tp}$ TPs identified by \TOOL and the $\mathit{fn}$ TPs missed by \TOOL; hence, recall is calculated as $\frac{\mathit{tp}}{(\mathit{tp} + \mathit{fn})}$.

\subsection{Methodology}
\label{subsec:methodology}

To answer the RQ, \TOOL and LintQ were applied to the same set of real-world Qiskit source code files.
We then compared the warnings generated by both tools, using the aforementioned annotated dataset, to assess their ability to identify actual quantum-specific problems in the code.

\subsection{Results}
\label{subsec:results}

Table~\ref{tab:results} 
presents the number of TPs, FPs, and NWs reported by each tool for each quantum-specific problem listed in Table~\ref{tab:lintq problems}.
We note that the number of TPs, FPs, and NWs obtained from \TOOL was determined through our manual inspection of the results it produced.
From our inspection, we found several warnings where LintQ and \TOOL identified the same problem occurrences but reported the corresponding code lines (i.e., locations) differently, as described below.

{
\renewcommand{\arraystretch}{0.85}
\begin{table}[t]
\caption{TPs, FPs, and NWs identified by LintQ and \TOOL.}
\label{tab:results}
\centering
\begin{tabularx}{\columnwidth}{lX@{}X@{}X@{}X@{}X@{}X}
\toprule
\multirow{2}{*}{Problem} & \multicolumn{3}{c}{LintQ} & \multicolumn{3}{c}{\TOOL} \\
\cmidrule(lr){2-4} \cmidrule(lr){5-7}
& \#TPs & \#FPs & \#NWs & \#TPs & \#FPs & \#NWs \\
\midrule
DoubleMeas & 18 & 4 & 3 & 17 & 3 & 3 \\
OpAfterMeas & 44 & 0 & 0 & 30 & 0 & 0 \\
MeasAllAbuse & 16 & 0 & 1 & 16 & 0 & 1 \\
CondWoMeas & 27 & 0 & 0 & 22 & 0 & 0 \\
ConstClasBit & 28 & 21 & 10 & 17 & 17 & 6 \\
InsuffClasReg & 24 & 22 & 21 & 9 & 3 & 7 \\
OversizedCircuit & 27 & 16 & 13 & 8 & 5 & 3 \\
GhostCompose & 7 & 0 & 3 & 4 & 0 & 0 \\
OpAfterOpt & 6 & 0 & 0 & 3 & 0 & 0 \\
OldIdenGate & 13 & 11 & 3 & 11 & 8 & 2 \\
\midrule
Sum & 210 & 74 & 54 & 137 & 36 & 22 \\
\bottomrule
\end{tabularx}
\end{table}
}

For example, LintQ and LintQ-LLM provide different but correct locations for the same occurrences of the ConstClasBit problem.
LintQ identifies occurrences of the ConstClasBit problem when a quantum circuit is created.
However, \TOOL identifies the same problem occurrence when a qubit is measured.
Recall from Section~\ref{sec:background} that the ConstClasBit problem occurs when a qubit is measured without undergoing any prior transformation. 
Hence, by definition, the warning locations (i.e., measurements in this case) identified by \TOOL are more aligned with the intended semantics of the problem.
Among the 137 TPs produced by \TOOL, 110 are found at the same locations identified by LintQ.
Of the remaining 27 TPs, 24 are found at locations involving measurements (21 TPs), operations (2 TPs), and register creation (1 TP), rather than at the quantum circuit creation locations where LintQ detects them.
The remaining three TPs occur when \TOOL and LintQ identify the same problem on different lines while referring to the same code segment (1 TP) and the double measurements on the same qubit (2 TPs).
This suggests that \TOOL provides a more precise localization of the problem compared to LintQ.

In terms of precision, \TOOL achieved a score of 70\%, which is in line with that of LintQ (62.6\%).
Although our results concern only the lines reported by both LintQ and \TOOL, and thus might be imprecise (e.g., TPs might be distributed differently in lines not annotated by LintQ), they are promising.

{
\renewcommand{\arraystretch}{0.8}
\begin{table*}[t]
\small
\caption{Recall for each problem detected by \TOOL.}
\label{tab:recall}
\centering
\begin{tabularx}{\textwidth}{@{}c@{\hspace{.1em}}c@{\hspace{.1em}}c@{\hspace{.1em}}c@{\hspace{.1em}}c@{\hspace{.1em}}c@{\hspace{.1em}}c@{\hspace{.1em}}c@{\hspace{.1em}}c@{\hspace{.1em}}c@{}}
\toprule
\rotatebox{15}{MeasAllAbuse} & \rotatebox{15}{DoubleMeas} & \rotatebox{15}{OldIdenGate} & \rotatebox{15}{CondWoMeas} & \rotatebox{15}{OpAfterMeas} & \rotatebox{15}{ConstClasBit} & \rotatebox{15}{GhostCompose} & \rotatebox{15}{OpAfterOpt} & \rotatebox{15}{InsuffClasReg} & \rotatebox{15}{OversizedCircuit} \\
\midrule
100\% & 94\% & 85\% & 81\% & 68\% & 61\% & 57\% & 50\% & 38\% & 30\% \\
\bottomrule
\end{tabularx}
\end{table*}
}

Table~\ref{tab:recall} presents the recall values for each problem detected by \TOOL.
Recall from Section~\ref{subsec:datasets} that the LintQ dataset was created by randomly sampling its own warnings; hence, computing LintQ's recall using this dataset is infeasible.
\TOOL detected most of the cases of
MeasAllAbuse, DoubleMeas, and OldIdenGate problems, with recall values above 85\%.
In contrast, \TOOL missed many 
InsuffClasReg and OversizedCircuit cases, with recall values of 38\% and 30\%, respectively.
By comparing 
these two groups of high-recall and low-recall problems, we found that \TOOL correctly detects TPs when the problems are related to API usage, such as measuring all qubits (MeasAllAbuse), individual measurements (DoubleMeas), and deprecated API usage (OldIdenGate).
However, \TOOL misses many TPs when problem identification requires complex analysis, such as data flow analysis on classical and quantum registers, corresponding to InsuffClasReg and OversizedCircuit, respectively.
The results suggest that future improvements should focus on enhancing \TOOL's capabilities for analyzing the usage flows of quantum and classical bits (i.e., data flows).
Overall, \TOOL achieved a recall score of 65\%; such a result is promising since it indicates that \TOOL can detect a high proportion of faults detected by means of static analysis. 
Our recall score is much higher than the 7.2\% recall score reported for LintQ~\cite{PaltenghiP24}; however, it is worth noting that we could not compare \TOOL on the same dataset used to compute LintQ recall (i.e., Bugs4Q~\cite{ZhaoMLZ23}). 
Indeed, the Bugs4Q dataset was made public before the training of any available LLM capable of code comprehension;
specifically, the oldest LLMs capable of code comprehension that are available are GPT-3.o and llama-2, both trained after the release of Bugs4Q.

\begin{tcolorbox}[left=2pt,right=2pt,top=0pt,bottom=0pt]
Our experimental results indicate that while neither LintQ nor \TOOL outperforms each other, \TOOL identifies warning locations that are more aligned with the semantics of the detected quantum-specific problems.
In addition, \TOOL performs well on API-related problems but struggles with complex problems requiring data flow analysis.
\end{tcolorbox}

\subsection{Threats to Validity}
\label{subsec:threats}

The main threats to the validity of our results is the possible bias introduced by the annotation dataset curated  by the authors of the LintQ paper.
Since the dataset includes \textit{samples} of warnings obtained from LintQ, it naturally cannot represent the entire population of warnings produced by \TOOL.
Nevertheless, we opted to use the annotated dataset, since it provides a common ground for comparison. For a fair comparison, further studies are needed with a new set of annotations that are free from bias toward either LintQ or \TOOL, such as those created independently from both tools.

To prevent any threats to validity caused by confounding factors, as discussed in Section~\ref{subsec:prompt engineering}, we selected the GPT-3.5 Turbo model, as its knowledge cutoff date is September 1, 2021, which is prior to the introduction of LintQ.
However, given the rapid advancements in the field of LLMs, applying newer models with improved reasoning capabilities and better understanding of source code may yield different results.  
Hence, future studies should consider evaluating \TOOL using more recent LLMs, such as GPT-o1~\cite{OpenAI2025o1} or beyond, to assess their capabilities in linting quantum programs. \section{Discussion}
\label{sec:discussion}

In this section, we discuss LintQ and \TOOL based on our experience of reproducing LintQ and developing \TOOL.
From these experiences, we have identified their distinct strengths and limitations, as well as potential research opportunities.

\textbf{Quantum programming frameworks.}
Although LintQ introduces abstractions for quantum programming constructs, it currently remains applicable only to Qiskit programs.
Extending LintQ to support other quantum programming frameworks, such as Cirq~\cite{cirq2025} and PennyLane~\cite{Bergholm2018}, requires manually mapping their APIs to the abstractions.
It may also necessitate modifying the abstractions and redefining queries to detect quantum-specific problems.
In contrast, \TOOL is easily adaptable to various quantum programming frameworks.
The only component requiring significant modification is the few-shot learning component, as it depends on framework-specific examples.

\textbf{Problem detection techniques.}
LintQ relies on the query-based problem detection scheme provided by CodeQL, making it capable of deterministically detecting problems.
However, detecting new quantum-specific problems---an expected challenge due to the advancements in quantum computing---requires manually crafting precise queries, which demands in-depth knowledge of both CodeQL and the target quantum programming framework.
In contrast, \TOOL leverages LLM-based analysis, allowing it to more easily adapt to detecting new problems without requiring manually defined queries.
This is because engineers facing new problems inherently have concrete examples that can be used for few-shot learning. 

\textbf{Handling large files.}
\TOOL leverages the GPT model, which operates within a token limit to optimize efficiency in a multi-user environment where it is accessed simultaneously.
Unfortunately, \TOOL failed to analyze 21 files out of 262 (see Section~\ref{subsec:datasets}) due to the token limit of the chosen underlying GPT model (i.e., \num{16385} tokens); such limitation, which is absent in LintQ, could be overcome either by using models with a higher token limit or by leveraging slicing techniques.

\textbf{Potential research avenues.}
A promising direction for future research is the development of \textit{hybrid} linting approaches that integrate the strengths of both tools.
For example, the outputs of LintQ and \TOOL could be cross-referenced to prioritize the warning identified by both.
In addition, the context-specific explanations provided by \TOOL could complement LintQ, helping developers better understand the identified warnings.
For example, \TOOL provides an explanation for an identified warning as follows: ``\textit{The QuantumCircuit `qc' is created with both a QuantumRegister `qr' and a ClassicalRegister `cr', but only one qubit is initialized in `qr'. This leads to an OversizedCircuit issue where resources are wasted on unused qubits}''.
Such explanations produced by \TOOL are significantly more informative compared to those produced by LintQ.
Building on this capability, interactive analysis powered by LLMs could further enhance the developer experience.
By engaging in a chatbot-based conversational feedback loop, developers can ask follow-up questions, seek clarification, or request code improvement suggestions directly within their development environment.
Last, LLMs could be used to automatically derive LintQ parsers for new quantum-programming languages, while \TOOL could support the identification of problems for which a LintQ analysis has not been implemented yet.

Our study restricted the comparison between LintQ and \TOOL to the analysis of Qiskit programs.
To derive more generalizable findings, further research is needed to evaluate these tools by applying them to other quantum programming frameworks.
In addition, conducting user-involved case studies is essential to assess the practical usefulness of these linting tools in real-world quantum software development. \section{Related Work}
\label{sec:related}

This section discuses the most pertinent research strands related to \TOOL: (1)~static analysis techniques for detecting quantum-specific programming problems, and (2)~LLM-assisted code analysis.

\textbf{Quantum-specific static analysis}.
Static analysis has been widely studied and applied for classical software programs; recently it has gained attention for quantum programs~\cite{ChenCCSA23,ZhaoWLZ23,KaulKB23,PaltenghiP24}.
Chen et al.~\cite{ChenCCSA23} identified eight quantum-specific code smells (referred to as ``problems'' in our context) from the Cirq best practices~\cite{CirqBestPractices}, validated them through a developer survey, and developed QSmell, a tool for detecting these smells.
QSmell employs both dynamic and static analysis techniques, leveraging execution details for the former and abstract syntax trees (ASTs) for the latter.
Regarding static analysis, QSmell identifies smells related to the use of non-parameterized circuits and the alignment between logical and physical qubits.
Zhao et al.~\cite{ZhaoWLZ23} introduced QChecker, a static analysis tool for detecting bugs (also referred to as ``problems'' in this paper) in quantum programs written in Qiskit.
QChecker identifies bugs based on predefined patterns in ASTs, derived from real-world quantum bugs~\cite{ZhaoMLZ23}.
It includes eight bug pattern detectors  covering both syntactic bugs and faulty logic in certain quantum-related operations.
Kaul et al.~\cite{KaulKB23} proposed QCPG, an extension of the Code Property Graph (CPG), which is a language-independent graph model  combining multiple representations---the AST, Program-Dependence Graph (PDG), and Control-Flow Graph (CFG)---for static code analysis.
QCPG represents quantum programs as graphs capturing quantum-specific language constructs, including quantum circuits, gates, qubits, and measurements.
Leveraging QCPG, Kaul et al.~\cite{KaulKB23} employed several graph queries to detect patterns corresponding to potential programming problems in Qiskit and OpenQASM~\cite{cross2022} code.
In contrast to \TOOL, these prior approaches rely on predefined rules, patterns, or graph constructs, which require considerable manual effort to update when quantum programming practices evolve.

\textbf{LLM-assisted code analysis}.
The use of LLMs for code analysis has attracted considerable interest  in the context of classical (non-quantum) software development~\cite{sun2024,wen2024,li2024,copilot,codewhisperer}.
For example, GitHub Copilot~\cite{copilot} provides code review capabilities that help developers identify potential problems and make improvements more efficiently.
Amazon CodeWhisperer~\cite{codewhisperer} suggests best coding practices and uncovers potential security vulnerabilities.
Despite these advancements, their application to quantum software remains limited.
\TOOL is the first known effort to leverage LLMs for quantum-specific linting.
We believe our work opens new opportunities for automated, adaptable, and developer-friendly quality assurance in quantum software development.

 \section{Conclusion}
\label{sec:conclusion}

As quantum computing continues to evolve, ensuring software quality remains a critical challenge.
In this paper, we explored the feasibility and potential of LLMs in quantum program analysis, highlighting the need for further advancements in linting techniques.
Specifically, we introduced \TOOL, an LLM-based linting tool for detecting quantum-specific programming problems, and conducted a comparative analysis with LintQ, a state-of-the-art query-based quantum-specific linting tool.
Our experiment results indicate that \TOOL is capable of detecting quantum-specific programming problems and, in some cases, provides better problem localization and more intuitive explanations than LintQ.
However, \TOOL did not outperform LintQ in detecting complex problems that require sophisticated static analysis techniques to track classical and quantum bit usage flows.
Based on these findings, we discussed the potential for a hybrid approach that integrates static analysis with LLM-powered analysis capabilities.
Such an approach could improve detection accuracy, offer developers context-aware explanations for detected problems, and  provide recommendations.
 
\section*{Acknowledgment}
We thank Matteo Paltenghi for the help with LintQ.

\bibliographystyle{IEEEtran}

\begin{thebibliography}{10}
\providecommand{\url}[1]{#1}
\csname url@samestyle\endcsname
\providecommand{\newblock}{\relax}
\providecommand{\bibinfo}[2]{#2}
\providecommand{\BIBentrySTDinterwordspacing}{\spaceskip=0pt\relax}
\providecommand{\BIBentryALTinterwordstretchfactor}{4}
\providecommand{\BIBentryALTinterwordspacing}{\spaceskip=\fontdimen2\font plus
\BIBentryALTinterwordstretchfactor\fontdimen3\font minus
  \fontdimen4\font\relax}
\providecommand{\BIBforeignlanguage}[2]{{\expandafter\ifx\csname l@#1\endcsname\relax
\typeout{** WARNING: IEEEtran.bst: No hyphenation pattern has been}\typeout{** loaded for the language `#1'. Using the pattern for}\typeout{** the default language instead.}\else
\language=\csname l@#1\endcsname
\fi
#2}}
\providecommand{\BIBdecl}{\relax}
\BIBdecl

\bibitem{aspuru2005}
A.~Aspuru-Guzik, A.~D. Dutoi, P.~J. Love, and M.~Head-Gordon, ``Simulated
  quantum computation of molecular energies,'' \emph{Science}, vol. 309, no.
  5741, pp. 1704--1707, 2005.

\bibitem{mcardle2020}
S.~McArdle, S.~Endo, A.~Aspuru-Guzik, S.~C. Benjamin, and X.~Yuan, ``Quantum
  computational chemistry,'' \emph{Reviews of Modern Physics}, vol.~92, no.~1,
  p. 015003, 2020.

\bibitem{ekert1991}
A.~K. Ekert, ``Quantum cryptography based on bell’s theorem,'' \emph{Physical
  review letters}, vol.~67, no.~6, p. 661, 1991.

\bibitem{portmann2022}
C.~Portmann and R.~Renner, ``Security in quantum cryptography,'' \emph{Reviews
  of Modern Physics}, vol.~94, no.~2, p. 025008, 2022.

\bibitem{harrow2009}
A.~W. Harrow, A.~Hassidim, and S.~Lloyd, ``Quantum algorithm for linear systems
  of equations,'' \emph{Physical review letters}, vol. 103, no.~15, p. 150502,
  2009.

\bibitem{ebadi2022q}
S.~Ebadi, A.~Keesling, M.~Cain, T.~T. Wang, H.~Levine, D.~Bluvstein,
  G.~Semeghini, A.~Omran, J.-G. Liu, R.~Samajdar \emph{et~al.}, ``Quantum
  optimization of maximum independent set using rydberg atom arrays,''
  \emph{Science}, vol. 376, no. 6598, pp. 1209--1215, 2022.

\bibitem{Nielson1999}
F.~Nielson, H.~R. Nielson, and C.~Hankin, \emph{Principles of program
  analysis}.\hskip 1em plus 0.5em minus 0.4em\relax Springer, 1999.

\bibitem{Dunsmuir1985}
M.~R.~M. Dunsmuir and G.~J. Davies, \emph{Program Analysis and
  Debugging}.\hskip 1em plus 0.5em minus 0.4em\relax Macmillan Education UK,
  1985, pp. 138--154.

\bibitem{ChenCCSA23}
Q.~Chen, R.~C{\^{a}}mara, J.~Campos, A.~Souto, and I.~Ahmed, ``The smelly
  eight: An empirical study on the prevalence of code smells in quantum
  computing,'' in \emph{Proceedings of the 45th {IEEE/ACM} International
  Conference on Software Engineering}, 2023, pp. 358--370.

\bibitem{ZhaoWLZ23}
P.~Zhao, X.~Wu, Z.~Li, and J.~Zhao, ``Qchecker: Detecting bugs in quantum
  programs via static analysis,'' in \emph{Proceedings of the 4th {IEEE/ACM}
  International Workshop on Quantum Software Engineering}, 2023, pp. 50--57.

\bibitem{KaulKB23}
M.~Kaul, A.~K{\"{u}}chler, and C.~Banse, ``A uniform representation of
  classical and quantum source code for static code analysis,'' in
  \emph{Proceedings of the 2023 {IEEE} International Conference on Quantum
  Computing and Engineering}, 2023, pp. 1013--1019.

\bibitem{PaltenghiP24}
M.~Paltenghi and M.~Pradel, ``Analyzing quantum programs with lintq: {A} static
  analysis framework for qiskit,'' \emph{Proceedings of the {ACM} on Software
  Engineering}, vol.~1, no. {FSE}, pp. 2144--2166, 2024.

\bibitem{Qiskit}
\BIBentryALTinterwordspacing
I.~Quantum, ``Qiskit: An open-source framework for quantum computing,'' 2025,
  accessed: March 10, 2025. [Online]. Available: \url{https://qiskit.org/}
\BIBentrySTDinterwordspacing

\bibitem{LLMtestingSurvey}
J.~Wang, Y.~Huang, C.~Chen, Z.~Liu, S.~Wang, and Q.~Wang, ``{Software Testing
  With Large Language Models: Survey, Landscape, and Vision},'' \emph{IEEE
  Transactions on Software Engineering}, vol.~50, no.~4, pp. 911--936, 2024.

\bibitem{AvgustinovMJS16}
P.~Avgustinov, O.~de~Moor, M.~P. Jones, and M.~Sch{\"{a}}fer, ``{QL:}
  object-oriented queries on relational data,'' in \emph{Proceedings of the
  30th European Conference on Object-Oriented Programming}, ser. Leibniz
  International Proceedings in Informatics (LIPIcs), vol.~56.\hskip 1em plus
  0.5em minus 0.4em\relax Schloss Dagstuhl - Leibniz-Zentrum f{\"{u}}r
  Informatik, 2016, pp. 2:1--2:25.

\bibitem{REPLICABILITY}
\BIBentryALTinterwordspacing
S.~Y. Shin, F.~Pastore, and D.~Bianculli, ``Replication package.'' [Online].
  Available: The distribution under a FOSS license is being reviewed by our legal team.
\BIBentrySTDinterwordspacing

\bibitem{paltenghi2023}
\BIBentryALTinterwordspacing
M.~Paltenghi and M.~Pradel, ``Analyzing quantum programs with lintq: A static
  analysis framework for qiskit,'' 2023. [Online]. Available:
  \url{https://arxiv.org/abs/2310.00718v1}
\BIBentrySTDinterwordspacing

\bibitem{PaltenghiLintQ}
\BIBentryALTinterwordspacing
------, ``Lintq: A static analysis framework for qiskit quantum programs,''
  2024, accessed: March 31, 2025. [Online]. Available:
  \url{https://github.com/sola-st/LintQ}
\BIBentrySTDinterwordspacing

\bibitem{OpenAIgpt35}
\BIBentryALTinterwordspacing
OpenAI, ``Gpt-3.5 turbo,'' 2025, accessed: March 31, 2025. [Online]. Available:
  \url{https://platform.openai.com/docs/models/gpt-3.5-turbo}
\BIBentrySTDinterwordspacing

\bibitem{burkov2025}
A.~Burkov, \emph{The Hundred-Page Language Models Book: hands-on with
  PyTorch}.\hskip 1em plus 0.5em minus 0.4em\relax True Positive Inc., 2025.

\bibitem{promptingguide}
\BIBentryALTinterwordspacing
DAIR.AI, ``Prompt engineering guide,'' 2025, accessed: March 20, 2025.
  [Online]. Available: \url{https://www.promptingguide.ai/}
\BIBentrySTDinterwordspacing

\bibitem{BrownMRSKDNSSAA20}
T.~B. Brown, B.~Mann, N.~Ryder, M.~Subbiah, J.~Kaplan, P.~Dhariwal,
  A.~Neelakantan, P.~Shyam, G.~Sastry, A.~Askell, S.~Agarwal,
  A.~Herbert{-}Voss, G.~Krueger, T.~Henighan, R.~Child, A.~Ramesh, D.~M.
  Ziegler, J.~Wu, C.~Winter, C.~Hesse, M.~Chen, E.~Sigler, M.~Litwin, S.~Gray,
  B.~Chess, J.~Clark, C.~Berner, S.~McCandlish, A.~Radford, I.~Sutskever, and
  D.~Amodei, ``Language models are few-shot learners,'' in \emph{Proceedings of
  the 2020 Annual Conference on Neural Information Processing Systems: Advances
  in Neural Information Processing Systems}.\hskip 1em plus 0.5em minus
  0.4em\relax Curran Associates, Inc., 2020, pp. 1877--1901.

\bibitem{ZhaoMLZ23}
P.~Zhao, Z.~Miao, S.~Lan, and J.~Zhao, ``Bugs4q: {A} benchmark of existing bugs
  to enable controlled testing and debugging studies for quantum programs,''
  \emph{Journal of Systems and Software}, vol. 205, p. 111805, 2023.

\bibitem{OpenAI2025o1}
\BIBentryALTinterwordspacing
OpenAI, ``Introducing openai o1,'' 2025, accessed: March 6, 2025. [Online].
  Available: \url{https://openai.com/o1/}
\BIBentrySTDinterwordspacing

\bibitem{cirq2025}
\BIBentryALTinterwordspacing
G.~Q. AI, ``Cirq: A python framework for near-term quantum computing,'' 2025,
  accessed: March 6, 2025. [Online]. Available:
  \url{https://quantumai.google/cirq}
\BIBentrySTDinterwordspacing

\bibitem{Bergholm2018}
V.~Bergholm, J.~A. Izaac, M.~Schuld, C.~Gogolin, and N.~Killoran, ``Pennylane:
  Automatic differentiation of hybrid quantum-classical computations,''
  \emph{CoRR}, vol. abs/1811.04968, 2018.

\bibitem{CirqBestPractices}
\BIBentryALTinterwordspacing
G.~Q. AI, ``Cirq best practices,'' 2025, accessed: March 10, 2025. [Online].
  Available: \url{https://quantumai.google/cirq/google/best_practices}
\BIBentrySTDinterwordspacing

\bibitem{cross2022}
A.~Cross, A.~Javadi-Abhari, T.~Alexander, N.~De~Beaudrap, L.~S. Bishop,
  S.~Heidel, C.~A. Ryan, P.~Sivarajah, J.~Smolin, J.~M. Gambetta \emph{et~al.},
  ``Openqasm 3: A broader and deeper quantum assembly language,'' \emph{ACM
  Transactions on Quantum Computing}, vol.~3, no.~3, pp. 1--50, 2022.

\bibitem{sun2024}
Y.~Sun, D.~Wu, Y.~Xue, H.~Liu, H.~Wang, Z.~Xu, X.~Xie, and Y.~Liu, ``Gptscan:
  Detecting logic vulnerabilities in smart contracts by combining gpt with
  program analysis,'' in \emph{Proceedings of the IEEE/ACM 46th International
  Conference on Software Engineering}, 2024, pp. 1--13.

\bibitem{wen2024}
C.~Wen, J.~Cao, J.~Su, Z.~Xu, S.~Qin, M.~He, H.~Li, S.-C. Cheung, and C.~Tian,
  ``Enchanting program specification synthesis by large language models using
  static analysis and program verification,'' in \emph{Proceedings of the 2024
  International Conference on Computer Aided Verification}, 2024, pp. 302--328.

\bibitem{li2024}
H.~Li, Y.~Hao, Y.~Zhai, and Z.~Qian, ``Enhancing static analysis for practical
  bug detection: An llm-integrated approach,'' \emph{Proceedings of the ACM on
  Programming Languages}, vol.~8, no. OOPSLA1, pp. 474--499, 2024.

\bibitem{copilot}
\BIBentryALTinterwordspacing
{GitHub}, ``{GitHub Copilot},'' 2025, accessed: March 10, 2025. [Online].
  Available: \url{https://github.com/features/copilot/}
\BIBentrySTDinterwordspacing

\bibitem{codewhisperer}
\BIBentryALTinterwordspacing
{Amazon}, ``{Amazon CodeWhisperer},'' 2025, accessed: March 10, 2025. [Online].
  Available: \url{https://aws.amazon.com/codewhisperer}
\BIBentrySTDinterwordspacing

\end{thebibliography}

\end{document}